\documentstyle[preprint,aps,epsf]{revtex}
\tightenlines 
\begin{document}
\draft

\title{Evidence for hadronic deconfinement in
$\bar{p}$-$p$ collisions at 1.8 TeV}

\author{T. Alexopoulos,$^{(1\ast)}$
E. W. Anderson,$^{(2)}$ 
A. T. Bujak,$^{(3)}$
D. D. Carmony$^{(3)}$
A. R. Erwin,$^{(1)}$ 
L.~J.~Gutay,$^{(3)}$ 
A.~S.~Hirsch,$^{(3)}$
K. S. ~Nelson,$^{(1**)}$
N.~T.~Porile,$^{(4)}$ 
S. H. Oh,$^{(6)}$
R.~P.~Scharenberg,$^{(3)}$ 
B.~K.~Srivastava,$^{(4)}$
B.~C.~Stringfellow,$^{(3)}$ 
F.~Turkot,$^{(7)}$ 
J.~Warchol,$^{(5)}$
W. D.~Walker$^{(6)}$ } 

\address{$^{(1)}$ Department of Physics, University of Wisconsin, Madison,
WI 53706} 
\address{$^{(2)}$ Department of Physics, Iowa State University, Ames, Iowa
50011} 
\address{$^{(3)}$ Department of Physics, Purdue University, West Lafayette, Indiana 47907}
\address{$^{(4)}$ Department of Chemistry, Purdue University, West
Lafayette, Indiana 47907}
\address{$^{(5)}$ Department of Physics, University of Notre Dame, Notre Dame,
Indiana 46556 } 
\address{$^{(6)}$ Department of Physics, Duke University, Durham, North Carolina
27706 }
\address{$^{(7)}$ Fermi National Accelerator Laboratory, Batavia, Illinois 60510 }

\date{\today}

\maketitle

\begin{abstract}
\vspace*{-.5in}

We have measured deconfined hadronic volumes, $4.4 < V < 13.0$ fm$^{3}$, 
produced by a one dimensional (1D) expansion. These volumes 
 are directly proportional to the
charged particle pseudorapidity
densities  $6.75 < dN_{c}/d\eta < 20.2$. The hadronization
temperature is $T = 179.5 \pm 5$ (syst) MeV. 
Using Bjorken's 1D model,
the hadronization energy density
is $\epsilon_{F} = 1.10 \pm 0.26$ (stat) GeV/fm$^{3}$
corresponding to an excitation of $24.8 \pm 6.2$ (stat) quark-gluon degrees
of freedom. 

\end{abstract}
\vspace*{12pt}


\newpage
\noindent
Corresponding Author:
\begin{minipage}[t]{4in}
{Prof. Rolf P. Scharenberg\\*[-8pt]
Department of Physics 1396\\*[-8pt]
Purdue University\\*[-8pt]
West Lafayette, IN 47907-1396\\
phone: (765) 494-5393\\*[-8pt]
fax: (765) 494-0706\\*[-8pt]
email: schrnbrg@physics.purdue.edu}
\end{minipage}

\vspace*{24pt}

The observation of high total multiplicity, high transverse energy, non jet,
isotropic events\cite{PL123B} led Van Hove\cite{hove82} and
Bjorken\cite{FL82} to conclude that high energy density events are produced
in high energy $\bar{p}$-$p$ collisions\cite{mclerran86}.  
These events have a far
greater cross section than jet production. In  these events the transverse
energy is proportional to the number of low transverse momentum particles.
This basic correspondence can be explored over a wide range of the charged
particle 
pseudorapidity density $dN_{c}/d\eta$ in $\bar{p}$-$p$ collisions at 
center of mass energy $\sqrt{s} = 1.8$ TeV.  
The various measurements from  the
Fermilab  quark-gluon plasma search 
experiment E-735 have already been published.  
In this letter, we  present for the first time a  coherent picture based
on the relationship between the volume $V$, 
temperature $T$,  energy density $\epsilon_{F}$, and  
pions per fm$^{3}$ $n_{\pi}$ emitted from $V$. 
 Spectra of identified particles, $\pi$,
$K$, $\varphi$, $p$, $\bar{p}$, $\Lambda^{o}$ $\overline{\Lambda^{o}}$,
$\Xi^{-}$, $\overline{\Xi^{-}}$
are used to extract  the $V$,  $\epsilon_{F}$, and $n_{F}$ values and to determine the 
strange quark content and relative yields of the hadrons.

Previously the various individual measurements did not provide an overall
understanding of these $\bar{p}$-$p$ collisions. Prompted by the new
analysis of the initial collision, we have developed a self-consistent
picture of hadronic deconfinement. This letter discusses:
{\bf (1)} The role of parton-parton (gluon) scattering;
{\bf (2)} The volume at decoupling, resulting 
from the one dimensional longitudinal expansion;
{\bf (3)}  The number of pions per fm$^{3}$ emitted by the source;
{\bf (4)} The hadronization temperature of the source;
{\bf (5)} The hadronization energy density of the source;
{\bf (6)} The number of quark-gluon degrees of freedom in the source;
{\bf (7)} The deconfined volumes and plasma lifetimes, estimates of initial 
energy
densities and temperatures.

Experiment E-735 \cite{FL83} was  located at the C$\emptyset$ interaction 
region of
the Fermi National Accelerator Laboratory (FNAL).  The
$\bar{p}$-$p$ interaction region was surrounded by a cylindrical drift chamber
which in turn was  covered by a single layer hodoscope including endcaps.
This system measured the total charged particle multiplicity 
$10 < N_{c} < 200$ in the pseudorapidity range  $ |\eta | < 3.25$. 
A magnetic spectrometer with tracking chambers and time of flight 
counters, provided
particle identified momenta spectra in the range $0.1 < p_{t} < 1.5$ GeV/c.
The spectrometer covered $-0.37 < \eta < + 1.00$ with $\Delta \varphi
\sim 20^{\circ}$ ( $\varphi$ is the azimuthal angle around the beam direction
). 

{\bf (1)} 
Recently the  E-735 collaboration has analyzed the
charged particle multiplicity distributions arising from $p$-$p$  and
$\bar{p}$-$p$ collisions over a range of center of mass energies 0.06
$\leq \sqrt{s} \leq 1.8$ TeV\cite{PLB435}.  Results at 1.8 TeV support
the presence of double $(\sigma_{2})$ and triple $(\sigma_{3})$ parton
interactions. These processes increase the non-single diffraction cross
section (NSD) from $\sim$ 32 mb at $\sqrt{s} = 0.06$ TeV to $\sim $ 48
mb at $\sqrt{s} = 1.8$ TeV.
The variation of the double encounter and triple encounter cross
sections $\sigma_{2}$ and $\sigma_{3}$ with center of mass energy
$\sqrt{s}$ is shown in Fig.\ 1.
\begin{figure}
\epsfxsize=8.5cm
\centerline{\epsfbox{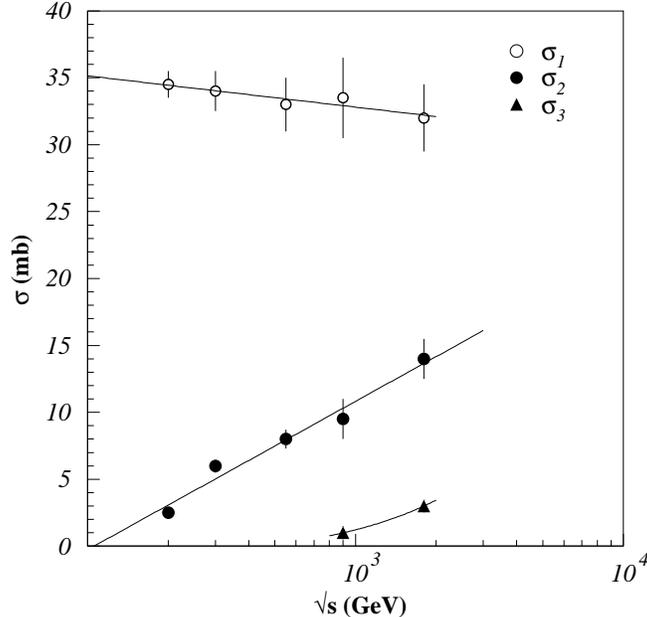}}
\vspace*{0.003in}
\caption{ Comparison of the cross sections for single, double, and triple
 encounter collisions which increase  $\rm \sigma_{NSD}$ 
above 32 mb as a function of $\sqrt{s}$.
}
\label{fig.1}
\end{figure}

The multiplicity distribution is made up of three contributions 
corresponding to single, double and triple parton-parton collisions. 
Our work on multiparton interactions shows that the increase in the
$p$-$p$ inelastic cross section with energy is nearly completely accounted
for by the increase in multiparton interactions. 
Previously this increase in the $p$-$p$ inelastic cross section was
ascribed to copious minijet production \cite{wang97}.
As the energy is increased, a decreasing fraction of the center of mass
energy appears in the NSD part of the inelastic 
cross section. This may be due to the decrease of 
the Feynman $x$ of the partons involved in these collisions. It
 is thus likely that gluons become more involved with increasing energy
leading to rapid thermalization \cite{muller99}.

{\bf (2)} To measure the hadronization volume $V$, pion 
 HBT (Hanbury Brown, Twiss) correlation
measurements were made as a function of both
$\vec{P}_{\pi\pi} = \vec{p}_{1} + \vec{p}_{2}$ the total momentum of
the pion pair and of ~$dN_{c}/d\eta$. The
$\vec{P}_{\pi\pi}$ momentum dependent results are shown in 
Table~I \cite{PRD48a}. 
$R_{G}$ is the Gaussian radius parallel to the beam, $\tau$ the
Gaussian lifetime, and $\lambda$  the chaoticity parameter. The
lifetime  $\tau$ can be viewed as a measure of the radius perpendicular
to the beam \cite{shuryak73}.
The increase of $R_{G}$ and $\tau$ with decreasing 
$P_{\pi\pi}$  is the characteristic signature for the expansion
of the pion source \cite{weiner}.
The dependence of
$R_{G}$ and $\tau$ on $dN_{c}/d\eta$ is shown in Table II \cite{PRD48a}. 
A clear increase
of $R_{G}$ with $dN_{c}/d\eta$ is evident.
The dependence of $R_{G}$ and $\tau$ on $P_{\pi\pi}$ and
$dN_{c}/d\eta$ is consistent with a one dimensional~(1D) longitudinal  
expansion of the
pion source. 
The effect of a 1D expansion on the Bose Einstein correlation
has been calculated for a massless relativistic ideal gas\cite{schlei92}.
 This calculation provides correction factors $\ell_{R}$ and $\ell_{\tau}$
to our values of  $R_{G}$ and $\tau$ obtained from the HBT analysis.
Both $\ell_{R}$ and $\ell_{\tau}$ are a function of $P_{\pi\pi}$ and
$\Delta \eta$, where
$\Delta \eta$ is the spectrometer aperture.
The cylindrical volume $V$ of the pion source is
$V = \pi 
(\ell_{\tau} \tau)^{2} \;
2 \; \ell_{R} R_{G}$
where $R_{G}$ varies with $dN_{c}/d\eta$ and $\ell_{\tau}\tau$ reaches an
asymptotic value for the larger $dN_{c}/d\eta$ values.
From our data $R_{G} = e + h$ $ dN_{c}/d\eta$ where 
$e = (0.0788 \pm 0.013)$ fm and 
$h = (0.0730 \pm 0.011)$ fm and $\chi^{2}/NDF$ = 3.09/4.00 as shown
in Fig.~2.
We neglect $e$ 
since $h \; dN_{c}/d\eta$ is  6 to 20 times larger than $e$. The
cylindrical volume becomes,
\begin{eqnarray}
\everymath={\displaystyle}       
V = \pi \ell_{\tau}^{2}  \tau^{2}  \;
2 \; \ell_{R} h \; dN_{c}/d\eta
\end{eqnarray}
\begin{figure}
\epsfxsize=8.5cm
\centerline{\epsfbox{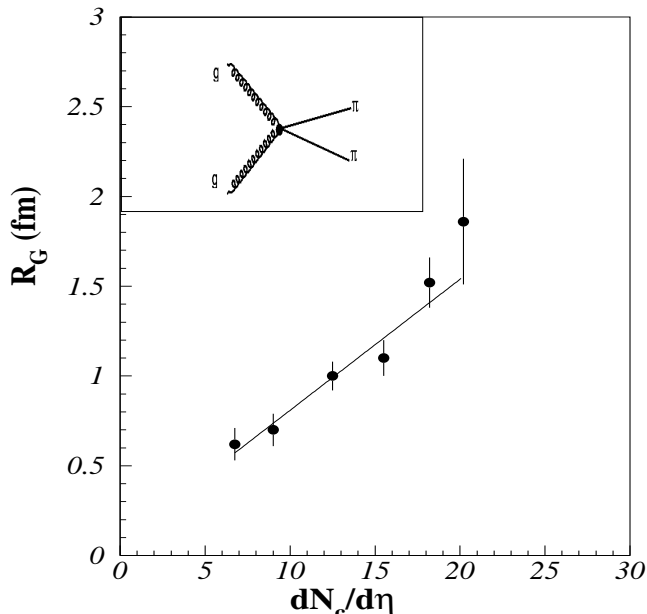}}
\vspace*{0.003in}
\caption{Dependence of the Gaussian radius $R_{G}$ on $dN_{c}/d\eta$.
The gluon diagram indicates that two  gluons are required to
form two pions.}
\label{fig.2}
\end{figure}

The largest measured value of $\tau = 0.95$ fm is used to evaluate $V$. We
estimate the $\ell$ factors using 
the extrapolation procedure ($P_{\pi\pi}$$\rightarrow$0) outlined in 
Ref.~\cite{schlei92}. ( See in particular Fig.4 and Eq.(7) in 
Ref~\cite{schlei92}).
For our $P_{\pi\pi}$ values and a 1D expansion, $\ell_{\tau} = 1$
independent of $P_{\pi\pi}$.
For the data in Table II $\ell_{R}= 1.56$. 
 Thus V = (0.645 $\pm $ 0.130)$dN_{c}/d\eta$ fm$^{3}$ and the range of V is 
4.4$\pm0.9<$ V$<$ 13.0$\pm2.6$ fm$^{3}$ for 6.75 $<$ $dN_{c}/d\eta < $ 20.2.  

{\bf (3)} We assume that for $dN_{c}/d\eta > 6.75$  the system is
above the deconfinement transition.
 The hot thermalized system expands, cools and then hadronizes.
We attribute all of the measured volume to the expansion before hadronization.
We neglect the subsequent expansion of the hadronic phase.
Following Bjorken's derivation,
we further
assume that hydrodynamics  of a massless relativistic ideal gas 
can describe the 1D expansion and that the
observed number of  pions/fm$^{3}$ are 
{\it proportional to the entropy density $s$ at
hadronization}. To estimate the pions/fm$^{3}$ emitted by the source, 
the Bjorken 1D boost invariant equation becomes
\begin{eqnarray}
\everymath={\displaystyle}       
s \; \propto \;  n_{\pi} =
\frac{3/2 \;\; dN_{c}/d\eta}
     {A \;\; 2 \;\; {\cal T}    }
\end{eqnarray}
where $A$ is the transverse area and $\cal T$ is the proper time
\cite{bjorken83}. 
The
collisions occur at longitudinal coordinate $z = 0$ and time $t = 0$. 
Eq.~(2) describes an isentropic  expansion $s ({\cal T}) / s({\cal T}_{0}) =
{\cal T}_{0}/ {\cal T}$ and  
\begin{eqnarray}
\everymath={\displaystyle}       
{\cal T} = (t^{2} - z^{2})^{1/2}
\end{eqnarray}
where ${\cal T}_{0}$ is the initial proper time when thermalization has
occurred.
For a relativistic massless ideal gas above the phase transition the
maximum expansion velocity, responsible for most of the longitudinal expansion,
is likely to be the sound velocity,  
$v_{s}^{2} = 1/3$\cite{bjorken83}.
The expansion time $t = z/v_{s} = \ell_{R} \; R_{G}/v_{s}$ and
${\cal T} = (3 z^{2} - z^{2})^{1/2}  = \sqrt{2} \; z$. We note that   
${\cal T}_{f}$ is the proper time at hadronization. 
\begin{eqnarray}
\everymath={\displaystyle}       
{\cal T}_{f} = \sqrt{2} \ell_{R} R_{G} =
\sqrt{2} \; \ell_{R} h \; dN_{c}/d\eta
\end{eqnarray}
and Eqn.~(2) becomes
\begin{eqnarray}
\everymath={\displaystyle}       
n_{\pi}~ = ~
\frac{3/2 \; \; dN_{c}/d\eta \; \; 1/\sqrt{2}}
     { \pi \tau^{2} \;\; 2 \;\; \ell_{R} h \; dN_{c}/d\eta}~ =~  
\frac{3/2 \; \; \; 1/\sqrt{2}}
 {\pi \tau^{2} \;\; 2 \;\; \ell_{R} h} \;
\end{eqnarray}
where $1/\sqrt{2}$ is the effective $\Delta \eta$ slice. Thus $n_{\pi}$ is independent of $dN_{c}/d\eta$ and one obtains 

\begin{eqnarray}
\everymath={\displaystyle}       
n_{\pi} = 1.64 \pm 0.33 \;\; \mbox{(stat) pions fm}^{-3}
\end{eqnarray}

This $n_{\pi}$ value indicates that the deconfinement transition
occurs at a definite entropy density.
Since $s \propto n_{\pi}$ is constant 
we can directly evaluate $n_{\pi}$ using the total number of pions emitted 
divided by the total
volume for the data set in Table ~I. We 
choose the lowest $P_{\pi\pi}$ value where $\tau = 0.95$ fm, $R_{G}$ = 1.2 fm,
and $\ell_{R} =$ 1.43. Here the average pseudo-rapidity  density is 
$< dN_{c}/d\eta> = 14.4$ and Eqns.~(2) and (4) become

\begin{eqnarray}
\everymath={\displaystyle}       
n_{\pi} = \frac{3/2 \; \; dN_{c}/d\eta \; \; 1/\sqrt{2}}
     { \pi \tau^{2} \;2 \; \ell_{R}\; R_{G}} &=&
     1.57 \pm 0.25 \;\; \mbox{(stat) pions fm}^{-3}
\end{eqnarray}
which has a smaller statistical error than (6).

{\bf (4)} The negative particle  $p_{t}$ spectrum is used to measure the temperature.
A slope parameter $b^{-1}$ is obtained from a fit of the invariant
cross section $d^{2}N_{c}/d y \; d^{2}p_{t}$ 
to the function $A \exp (-bp_{t})$ for $0.15 \leq p_{t} \leq 0.45$
GeV/c \cite{PLB336}.  The $b^{-1}$ value is constant
to $\pm$ 1\% for $6.75  < dN_{c}/d\eta < 20.2$. 
Transverse flow has not been seen in $p$-$p$ reactions
at lower energies \cite{laasanen,bearden}.  In heavy ion reactions
the transverse flow is attributed to final state interactions 
of the hadrons which
presumably are not important in $\bar{p}$-$p$ collisions.  
The fact that $R_{G}$ increases by a factor
of three and $b^{-1}$ remains constant to $\pm$ 1\%, suggests that the
transverse flow is negligible.  The components
$\sigma_{2},\sigma_{3}$ in the NSD cross section indicate that the
parton-parton mean free paths are shorter in high energy collisions.
Since gluon-gluon interactions dominate in the initial encounters,
early thermalization $\sim$ 0.5 fm/c when $T \sim$ 200 MeV is
likely\cite{muller99}.  We interpret $b^{-1} = T = 179.5 \pm 5$
MeV (syst) as the hadronization temperature.
We neglect the expansion of the hadronic phase following
hadronization i.e. decoupling is associated with hadronization. 
 The systematic error estimate is based on possible kaon ($K_{s}^{0}$) 
misidentification 
in the negative particle spectrum at low $p_{t}$. We have not made a correction
for the effect of resonance decays on the negative particle $p_{t}$ spectrum.
We note that the negative particle temperature is significantly higher 
 than the
 temperature based on the spectra of idenitified pions which include 
 resonance particle decay pions~~~~(T$\simeq$168 MeV).
A  hadronization inverse slope parameter $T_{m}$ can be estimated
from our measurement of the relative yields of mesons and hyperons as shown
in Fig.~3, using all the events with $dN_{c}/d\eta > 6.75$.  The hadron
yield versus rest mass inverse slope parameters indicates 
$162 < T_{m} < 173$
MeV. Similar 
$T_{m} \sim 168 $ MeV values, 
based on thermal model analyses of hadron yield ratios, 
have been seen in high energy $\bar{p}$ p, p p, $e^{+} e^{-}$ and heavy ion
reactions \cite{heinz01}. This has  been interpreted  
 as evidence for a universal limiting temperature $T_{m}$ for hadrons, the Hagedorn temperature\cite{hagedorn80}.
\begin{figure}
\epsfxsize=8.5cm
\centerline{\epsfbox{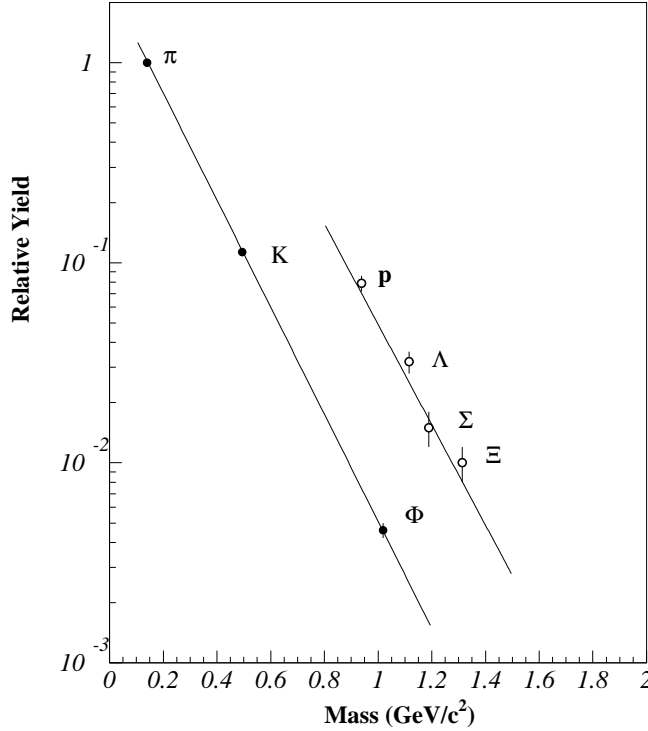}}
\vspace*{0.003in}
\caption{Relative meson and hyperon yields versus 
rest mass[19,20]. 
For the mesons, the inverse slope parameter  $T_{m}$ = 162$\pm$5 MeV, 
and  for the hyperons $T_{m}$ = 173$\pm$ 12 MeV.
}
\label{fig.3}
\end{figure}

{\bf (5)} We can also use the average  measured energies to estimate 
the hadronization energy density $\epsilon_{F}$ \cite{PRD48b}. 
Since $\epsilon_{F} = 3/4 \; T\; s_{F}$, 
~$\epsilon_{F}$ is also constant versus $dN_{c}/d\eta$ \cite{bjorken83}, 
\begin{eqnarray}
\everymath={\displaystyle}       
\epsilon_{F} =
\frac{ \sum_{h} F_{h} \times  (m_{h})_{\perp} \times 1 / \sqrt{2}}
     {\pi \tau^{2} \times 2 \times \ell_{R} \times h}
\end{eqnarray}
where $(m_{h})_{\perp} = (m_{h}^{2} + p_{t}^{2})^{1/2}$ is the average
  transverse
mass of hadron $h$, $F_{h}$ is a hadron abundance factor which also
accounts for the neutral hadrons of each species. We have determined 
$F_{h}$ for  $\pi, K, \varphi, p, n, \Lambda^{0}, \Xi$, etc. 
For  $\tau = 0.95$ fm, $\ell_{R}$ = 1.56, h=0.073, 
~ $\epsilon_{F}$ becomes
\begin{eqnarray}
\everymath={\displaystyle}       
\epsilon_{F} = 1.10 \pm 0.26 \; \mbox{(stat) GeV/fm}^{3}
\end{eqnarray}

{\bf (6)} We can estimate the average number $n_{c}$ of constituents  
in volume $V$ at  
temperature $T$, for a system without boundaries\cite{pathria72}
\begin{eqnarray}
\everymath={\displaystyle}       
n_{c} = V \; 
\frac{G(T) \; 1.202 \;\; (kT)^{3}}
     {\pi^{2} \;\; \hbar^{3} \;\; c^{3}}
\end{eqnarray}
where $G(T)$ are the number of degrees of freedom (DOF).
For a pion gas $G(T) = 3$, $V =$ 1 fm$^{3}$, 
and $T = 179.5$ MeV. 
The average  number  of pions (pion gas) {\em in the source} is
$n_{\pi} = 0.28$
pions/fm$^{3}$.
We observe 1.57 pions/fm$^{3}$, emitted from the source at temperature $T = 179.5$ MeV, which requires many more DOF.

For a quark-gluon plasma 
$ G(T) = G_{g}(T) + G_{q}(T) +
G_{\bar{q}}(T) = 16 + 21/2 \; (f)$ where $f$ are the
number of quark flavors \cite{bjorken83}. 
$G_{g}(T)$ are the gluon DOF;~$G_{q}(T)$, $G_{\bar{q}}(T)$ are the quark,
antiquark DOF.
We $\it assume$ that pion emission from the
source can be determined by the number of constituents in the source at
hadronization, that one
 pion is a quark antiquark $(q,\bar{q})$ pair and that 
two gluons $(2g)$
are required to produce two pions (see insert Fig.~2).
\begin{equation}
n_{\pi} =
n_{g} + (n_{q} + n_{\bar{q}})/2
\end{equation}

Our data indicates that $\sim$ 6\% strange quarks are present at
hadronization \cite{PRD48b,zphysc67}. Thus we use $f = 2$ to evaluate Eqn.~(10) where $V = 1$
fm$^{3}$ and,
\begin{equation}
n_{\pi} = (1 + 2 \times 21/64 ) G_{g} \times 16.1 \; T^{3} \; \mbox{(GeV)}
\end{equation}
where $G_{g}$ are
the effective number of gluon DOF. For $n_{\pi} = 1.57$/fm$^{3}$
and $T = 0.1795$ GeV,
we obtain $G_{g} = 10.18$. The total number of DOF are,
\begin{equation}
G(T) =  n_{g} + n_{q} + n_{\bar{q}} =
(1 + 21/16 ) G_{g} =
23.5 \pm 6  \; \mbox{DOF}
\end{equation}
nearly eight times the DOF for a pion gas.
A second method for estimating the DOF is to use the energy density and
temperature at hadronization.
For the isentropic expansion,
the energy $E$ in the volume $V$ at temperature $T$ is \cite{pathria72}
\begin{equation}
E = V  \;
\frac{G(T) \; \pi^{2} \; k^{4}}
     {30 \; \hbar^{3} \; c^{3}}
T^{4} \ . 
\end{equation}
For $\epsilon_{F} = 1.10 \pm 0.22$ (stat) GeV/fm$^{3}$
and $T= 179.5 \pm 5$ MeV, 
we find $G(T) = 24.8 \pm 6.2$ (stat) quark
gluon DOF, in good agreement with the DOF using the number of
constituents (Eq. 13).

{\bf (7)} Two Lorentz contracted nucleons collide at $t = 0$, $z = 0$ and the
thermalized  
constituents  are assumed to 
emerge at ${\cal T}_{0}$. Suppose we choose ${\cal T}_{0} = 1.0$ fm/c. 
For a given expansion velocity,
the data determines the hadronization proper time ${\cal T}_{f}$ and $1.09  <
{\cal T}_{f} < 3.25$ fm/c. 
For $6.75 < dN_{c}/d\eta < 20.2$,
the deconfined volumes $V$, determined by the data, range between $4.4 < V
< 13.0$ fm$^{3}$. For $dN_{c}/d\eta > 6.75$ and using $G(T)$ from Eq.~(14)

\begin{equation} 
\epsilon /T^{4} = \pi^{2}/30 ~G(T) = 8.15 \pm 2.0 (stat) 
\end{equation}
in general agreement with lattice gauge calculations
\cite{karsch}.  The ratio of the initial temperature $T_{i}$ to the final
$T_{f}$ is $T_{i}/T_{f} = ({\cal T}_{f}/{\cal T}_{0})^{1/3}$ and $185 < T_{i} <
266$ MeV. The ratio of the initial energy density $\epsilon_{i}$ to the
final energy density $\epsilon_{f}$ is $\epsilon_{i}/\epsilon_{f} 
= ({\cal T}_{f} /{\cal T}_{0})^{4/3}$ and  $\epsilon_{i}$ is
 $1.23 < \epsilon_{i} < 5.30$
GeV/fm$^{3}$ for $6.75 < dN_{c}/d\eta < 20.2$.
Note a different choice of ${\cal T}_{0}$ would change the $T_{i}$ and 
$\epsilon_{i}$
estimates.

In summary, the HBT analysis 
and the constant temperature versus  $dN_{c}/d\eta$ 
are consistant  with a model in which a pion source undergoes a 1D expansion
with total longitudinal dimension 2~$l_{R} R_{G}$ directly proportional 
to $dN_{c}/d\eta$.  
We have used the Bjorken 1D model to analyze
our data. We find that there is a unique hadronization entropy density and
temperature at which the pions are produced independent of
$dN_{c}/d\eta$.  We have used phase space estimates of the average
number of thermalized constituents in volume $V$ at temperature $T$ and
the measured energy density $\epsilon_{F}$ to compute the number of DOF
in the source.  
However, we note that reducing the average expansion velocity from $v^{2} =
1/3$ to $v^{2} = 1/5$ reduces the DOF estimate by 30\%. 
Then the  lower limit  for the DOF 
is  16.6$\pm$4.2, still 
substantially larger than the pion gas DOF of 3. This lower limit allows a more
conservative argument that quark-gluon constituents are present in the large  
deconfined volumes.  
Our estimate of
the number of DOF in the source (23.5$\pm$6, 24.8$\pm$6.2) is in general 
agreement
with those expected for a quark-gluon plasma.  
The $n_{\pi}$, $\epsilon_{F}$, and $T$ values 
characterize the quark-gluon to hadron thermal phase transition.
We expect that these   
hadronization conditions will be observed in
$p$-$p$ collisions at the  CERN
Large Hadron Collider where higher pseudorapidity
density $dN_{c}/d\eta$ values will produce even larger deconfined volumes 
and longer plasma lifetimes.

\vspace*{12pt}

We would like to acknowledge the important support for Fermi National
Laboratory  experiment E-735 by J. D. Bjorken and L. M. Lederman.
This work was supported in part by the United States Department of
Energy and the National Science Foundation.


\newpage

\begin{table}
\caption{ Fitted values of radius $R_{G}$, lifetime $\tau$, and chaoticity
$\lambda$ in the Gaussian parameterization with respect to $q_{t}$ and
$q_{0}$. Values are a function of average two-pion total momentum
$P_{\pi\pi}$ or average two-pion transverse momentum $P_{t}$. The total
momentum interval containing the data is listed in column 1. Momentum is in
GeV/c. The errors are statistical. }
\begin{tabular}{c c c c c c}
$P_{\pi \pi}$ &
$\langle P_{\pi \pi}\rangle$ &
$R_{G}$ (fm) &
$\tau$ (fm) &
$\lambda$ &
$\langle P_{t} \rangle$ \\\hline
0.2-0.5 & 0.404 & 1.20$\pm$0.05 & 0.95$\pm$0.06 & 0.24$\pm$0.01 & 0.369 \\
0.2-0.7 & 0.503 & 1.05$\pm$0.08 & 0.71$\pm$0.05 & 0.25$\pm$0.01 & 0.462\\ 
0.5-1.0 & 0.708 & 0.80$\pm$0.07 & 0.67$\pm$0.07 & 0.23$\pm$0.02 & 0.650\\
0.7-1.2 & 0.900 & 0.60$\pm$0.06 & 0.64$\pm$0.05 & 0.26$\pm$0.03 & 0.832\\
0.9-1.7 & 1.175 & 0.58$\pm$0.06 & 0.53$\pm$0.07 & 0.26$\pm$0.02 & 1.087 \\
$>$1.0  & 1.403 & 0.48$\pm$0.06 & 0.45$\pm$0.05 & 0.21$\pm$0.02 & 1.285 \\
$>$1.2  & 1.600 & 0.43$\pm$0.06 & 0.41$\pm$0.06 & 0.23$\pm$0.02 & 1.479
\end{tabular}
\end{table}

\newpage

\begin{table}
\caption{Fitted values of the (longitudinal) radius $R_{G}$ (transverse)
lifetime $\tau$ and corrected chaoticity $\lambda$ in the Gaussian
parameterization with respect to $q_{t}$ and $q_{0}$.  Values are a function of the
average charged multiplicity per unit of pseudorapidity. Charged particle 
multiplicity intervals containing the data are listed in column 1. The
errors are statistical.} 
\begin{tabular}{c d c c c}
$N_{c}$ & 
$\langle dN_{c}/d\eta \rangle$ &
$R_{G} (fm)$ & 
$\tau$ (fm) &
$\lambda$ \\\hline
0-60    &  6.75 & 0.62 $\pm$ 0.09  & 0.53 $\pm$ 0.07 & 0.39 $\pm$ 0.05 \\
0-80    &  9.00 & 0.70 $\pm$ 0.09  & 0.65 $\pm$ 0.06 & 0.32 $\pm$ 0.03 \\
60-100  & 12.5 & 1.00 $\pm$ 0.08  & 0.86 $\pm$ 0.12  & 0.25 $\pm$ 0.01 \\
80-120  & 15.5 & 1.10 $\pm$ 0.10  & 0.89 $\pm$ 0.12 & 0.23 $\pm$ 0.01 \\
100-240 & 18.2 & 1.52 $\pm$ 0.14  & 0.99 $\pm$ 0.15 & 0.21 $\pm$ 0.02 \\
120-240 & 20.17 & 1.86 $\pm$ 0.35 & 0.88 $\pm$ 0.20 & 0.19 $\pm$ 0.03\\ 
\end{tabular}
\end{table}

\begin{references}
\bibitem[*]{}
Current Address: Department of Physics, National Technical University of
Athens, Athens, Greece. 

\bibitem[**]{}
Current Address: Department of Physics, University of Virginia,
Charlottesville, Virginia.


\bibitem{PL123B} 
UA1 collaboration, Phys. Lett. B {\bf 123} (1983) 115;
Physics Letters {\bf 107B} (1981) 320.

\bibitem{hove82}
L. Van Hove, Phys. Lett. B {\bf 118} (1982) 138.

\bibitem{FL82}
J. D. Bjorken, Fermilab Pub. 82/44-THY.

\bibitem{mclerran86}
L. McLerran, Rev. Mod. Phys. {\bf 58} (1986) 1021.

\bibitem{FL83}
E-735 proposal. Search for a deconfined quark-gluon plasma phase, Fermilab
P-735 (1983).

\bibitem{PLB435}
T. Alexopoulos, {\it et al.}, 
(E-735 collaboration)
Phys. Lett. B {\bf 435} (1998) 453;
W. D. Walker, private communication. 


\bibitem{wang97}
X. N. Wang, Phys. Rep. {\bf 280} (1977) 287.

\bibitem{muller99}
B. M\"{u}ller and A. Trayanov, Int. J. Mod. Phys. {\bf C5} (1994) 113;
W. P\"{o}schel and B. M\"{u}ller, Phys. Rev. D {\bf 60} (1999) 114505. 

\bibitem{PRD48a}
T. Alexopoulos, {\it et al.}, 
(E-735 collaboration) 
Phys. Rev. D {\bf 48} (1993) 1931.

\bibitem{shuryak73}
E. V. Shuryak, Phys. Lett. {\bf 44B} (1973) 387;
G. Cocconi, Phys. Lett. {\bf 49B} (1974) 459.

\bibitem{weiner}
R. M. Weiner, Physics Reports {\bf 127} (2000) 249.

\bibitem{schlei92}
B. R. Schlei, {\it et al.}, Phys. Lett. B {\bf 293} (1992) 275.

\bibitem{bjorken83}
J. D. Bjorken, Phys. Rev. D {\bf 27} (1983) 140.

\bibitem{PLB336}
T. Alexopoulos, {\it et al.}, 
(E-735 collaboration),
Phys. Lett. B {\bf 336} (1994) 599.

\bibitem{laasanen}
A. T. Laasanen, et al., Phys. Rev. Lett. {\bf 38} (1977) 1.

\bibitem{bearden}
I. Bearden, et al. (NA44 Collaboration). Phys. Rev. Lett. {\bf 78} (1977)
2080.

\bibitem{heinz01}
U. Heinz, Nucl. Phys.  {\bf 685} (2001) 414c.
\bibitem{hagedorn80}
R. Hagedorn and J. Rafelski , Phys. Lett. {\bf 97B} (1980) 136.
\bibitem{PRD48b}
T. Alexopoulos, {\it et al.}, 
(E-735 collaboration), 
Phys. Rev. D {\bf 46} (1992) 2773;
Phys. Rev. D {\bf 48} (1993) 984.

\bibitem{zphysc67}
T. Alexopoulos, {\it et al.}, 
(E-735 collaboration), Z. Phys. {\bf C67} (1995) 411.
\bibitem{pathria72}
R. K. Pathria, Statistical Mechanics, Pergamon Press, Ltd, 1972, p 187f. 


\bibitem{karsch}
F. Karsch, {\it et al.}, Phys. Lett. B {\bf 478} (2000) 249.

\end{references}
\end{document}